\documentclass[fleqn,twoside]{article}
\usepackage[headings]{espcrc2}
\usepackage{epsf,epsfig}
\readRCS $Id: espcrc2.tex,v 1.2 2004/02/24 11:22:11 spepping Exp $
\usepackage{graphicx}
\usepackage[figuresright]{rotating}

\def\beq{\begin{eqnarray}}
\def\eeq{\end{eqnarray}}
\def\bea{\begin{eqnarray}}
\def\eea{\end{eqnarray}}\def\be{\begin{equation}}
\def\ee{\end{equation}}

\newcommand{\AmS}{{\protect\the\textfont2
  A\kern-.1667em\lower.5ex\hbox{M}\kern-.125emS}}

\title{Rare B decays in a single Universal Extra Dimension scenario}

\author{F. De Fazio\address{Istituto Nazionale di Fisica
Nucleare, Sezione di Bari, Italy}}

\begin{document}

\begin{abstract}
Exclusive  rare  $B \to K^{(*)} \ell^+ \ell^-$,  $B \to K^{(*)}
\nu \bar \nu$ and $B \to K^* \gamma$ decays are studied within the
Applequist-Cheng-Dobrescu model,  an extension of the Standard
Model in presence of universal extra dimensions. In  the case of a
single universal extra dimension compactified on a circle of
radius R, we study the dependence of several observables on $1/R$,
and discuss  whether the hadronic uncertainty due to the form
factors obscures or not such a dependence. We find that, using
present data, it is possible in many cases to put a sensible lower
bound to  $1/R$, the most stringent one coming from $B \to K^*
\gamma$. \vspace{1pc}
\end{abstract}

\maketitle

\section{INTRODUCTION}

Among the various  new Physics scenarios, those with extra
dimensions are particularly attracting \cite{rev}. A special case
is
  the Appelquist-Cheng-Dobrescu (ACD)
  model \cite{Appelquist:2000nn} in which universal extra dimensions are
considered, which means that
 all the fields can
 propagate in all available dimensions.
  In the case of  a single  extra
dimension compactified on a circle of radius R, Tevatron run I
data  allow to put the bound $1/R \ge 300 $ GeV. Constraints can
be put   studying other processes, namely rare B decays induced by
$b \to s$ transition \cite{Agashe:2001xt} which are  induced at
loop level and hence suppressed in  the Standard Model (SM)
\cite{Hurth:2003vb}.

In  \cite{Buras:2002ej,Buras:2003mk} the effective Hamiltonian
inducing $b \to s$ decays was computed in the ACD model. Here, we
summarize the results obtained in \cite{Colangelo:2006vm} for
exclusive $b \to s$-induced modes.
 In this case, the uncertainty in the  form factors must be considered,
  since it can  overshadow the sensitivity to  $1/R$. Indeed, we find that
 computing the  branching
 ratios of $B \to K^{(*)} \ell^+  \ell^-$ and the forward-backward lepton asymmetry
 in $B \to K^* \ell^+  \ell^-$ for a representative set of form factors,  a bound can
 be put. Other interesting observables are the lepton polarization
 asymmetries in the case of the modes $B \to K^{(*)} \tau^+  \tau^-$.
 Finally,  we discuss how the branching ratio of
  $B \to K^* \gamma$ depends on   $1/R$,  from which we can
 establish the most stringent  bound on this parameter.

\section{THE ACD MODEL WITH A SINGLE UED}

The  ACD model \cite{Appelquist:2000nn}   is a  minimal extension
of the SM in $4+ \delta$ dimensions; we  consider
 $\delta=1$. The fifth dimension $x_5=y$ is compactified
to the orbifold $S^1/Z_2$, i.e. on a circle of radius R
  and runs
from 0 to $ 2 \pi R$ with   $y=0, y=\pi R$  fixed points of the
orbifold. Hence a field $F(x,y)$ ($x$ denoting the  usual 3+1
coordinates) would be a
 periodic function of $y$,  and  it could be expressed as
 $\displaystyle{F(x,y)=\sum_{n=- \infty}^{n=+  \infty} F_n(x) e^{i\, n \cdot
 y/R}}$.
If $F$ is  a massless boson field, the KK modes $F_n$  obey the
equation
 $\displaystyle{\left(\partial^\mu
\partial_\mu+ n^2/ R^2 \right)F_n(x)=0 }$,
 $\mu=0,1,2,3$ so that, apart the zero mode, they
 behave in four dimensions as massive  particles with $m_n^2= (n/R)^2$.
Under the parity transformation $P_5: y \to -y$ fields having a
correspondent in
 the 4-d SM should be even, so that their zero mode in the  expansion is interpreted
as the ordinary SM field. On the other hand, fields having no
 SM partner should be odd,  so that  they do not
have zero modes.

Important features of the ACD model are: i)
 a single  additional free parameter with respect to the SM:
the compactification radius $R$; ii)  conservation of KK parity,
so that there is no tree-level contribution of KK modes in low
energy processes and no production of single KK excitation in
ordinary particle
 interactions.
A detailed description  of  this model is provided in
\cite{Buras:2002ej}.

\section{DECAYS $ B \to K^{(*)} \ell^+ \ell^-$ } \label{sec:modes}

In the Standard Model the effective $ \Delta B =-1$, $\Delta S =
1$ Hamiltonian governing  the  transition $b \to s \ell^+ \ell^-$
is $ H_W\,=\,4\,{G_F \over \sqrt{2}} V_{tb} V_{ts}^*
\sum_{i=1}^{10} C_i(\mu) O_i(\mu)$. \noindent $G_F$ is the Fermi
constant and $V_{ij}$ are elements of the
Cabibbo-Kobayashi-Maskawa mixing matrix; we neglect terms
proportional to $V_{ub} V_{us}^*$. $O_1$, $O_2$ are
current-current operators, $O_3,...,O_6$   QCD penguins, $O_7$,
$O_8$  magnetic penguins, $O_9$, $O_{10}$ semileptonic electroweak
penguins. We do not consider
 the contribution to $B \to K^{(*)} \ell^+ \ell^-$  with the  lepton pair  coming from
 $c{\bar c}$ resonances,  mainly due to  $O_1$, $O_2$.
We also neglect QCD penguins whose  coefficients are very small
compared to the others. Therefore, in the case of the modes $B \to
K^{(*)} \ell^+ \ell^-$, the relevant operators are: $O_7={e \over
16 \pi^2} m_b ({\bar s}_{L \alpha} \sigma^{\mu \nu}
     b_{R \alpha}) F_{\mu \nu} $,
$O_9={e^2 \over 16 \pi^2}  ({\bar s}_{L \alpha} \gamma^\mu
     b_{L \alpha}) \; {\bar \ell} \gamma_\mu \ell $, $
O_{10}={e^2 \over 16 \pi^2}  ({\bar s}_{L \alpha} \gamma^\mu
     b_{L \alpha}) \; {\bar \ell} \gamma_\mu \gamma_5 \ell
$. Their coefficients   have been computed at NNLO in the Standard
Model \cite{nnlo} and at
 NLO   for  the ACD model \cite{Buras:2002ej,Buras:2003mk}: we
 use these results in our study. No new operators  are found in
 ACD, while
  the   coefficients  are modified because particles not
present in  SM can contribute as intermediate states in loop
diagrams. As a consequence, they are expressed in terms of
functions $F(x_t,1/R)$, $x_t=m_t^2/M_W^2$,  generalizing the
corresponding SM functions $F_0(x_t)$ according to
$\displaystyle{F(x_t,1/R)=F_0(x_t)+\sum_{n=1}^{\infty}F_n(x_t,x_n)}$,
 where $x_n=m_n^2/M_W^2$
and $m_n=n/R$ \cite{Buras:2002ej,Buras:2003mk}.
 For large values of $1/R$ the SM phenomenology should be
  recovered, since the new states,  being more and more massive,  decouple from the low-energy theory.

The exclusive $B \to K^{(*)} \ell^+ \ell^-$ modes involve
  the matrix elements of the operators  in
the effective Hamiltonian  between the $B$ and  $K$ or $K^*$
mesons, for which we use the standard parametrization in terms of
form factors\footnote{See \cite{Colangelo:2006vm} for a detailed
discussion.}. We use two sets of  form factors:
 the first one (set A) obtained by  three-point QCD sum rules
based on the short-distance expansion \cite{Colangelo:1995jv}; the
second one (set B) obtained by light-cone QCD sum rules
\cite{Ball:2004rg}.  For both sets we include in the numerical
analysis the errors on the parameters.

 In Fig. 1 we
plot, for the two sets of form factors,  the branching fractions
relative to $B \to K^{(*)} \ell^+ \ell^-$ versus $1/R$
 and compare them with the experimental data provided by
BaBar \cite{Aubert:2004it}: $BR(B\to K \ell^+ \ell^- )=(3.4 \pm
0.7 \pm 0.3) \times 10^{-7}$, $BR(B \to K^* \ell^+ \ell^-)=(7.8
\pm^{1.9}_{1.7} \pm 1.2)\times 10^{-7}$, and Belle
\cite{Iwasaki:2005sy}: $BR(B\to K \ell^+ \ell^- )=(5.50
\pm^{0.75}_{0.70} \pm 0.27 \pm 0.02) \times 10^{-7}$, $BR(B \to
K^* \ell^+ \ell^-)=(16.5 \pm^{2.3}_{2.2} \pm 0.9 \pm 0.4)\times
10^{-7}$.

 Set B
 excludes  $1/R \leq 200$ GeV.
  Improved
data will resolve the  discrepancy between the  experiments and
increase the lower bound for $1/R$.

In the case of  $ B \to K^* \ell^+ \ell^-$ the  investigation of
the forward-backward asymmetry ${\cal A}_{fb}$ in the dilepton
angular distribution may also reveal effects beyond the SM. In
particular,
 in  SM,  due to the opposite sign of the
coefficients $C_7$ and $C_9$, ${\cal A}_{fb}$ has a zero the
position of which is almost independent of the model for  the form
factors \cite{Burdman:1998mk}. Let  $\theta_\ell$ be the angle
between the $\ell^+$ direction and the $B$ direction in the rest
frame of the lepton pair (we consider  massless leptons). We
define ($z={\rm cos}\theta_\ell$):
\begin{equation}
{\cal A}_{fb} (q^2)=\displaystyle {\displaystyle \int_0^1{d^2
\Gamma \over dq^2 d z}dz -\int^0_{-1}{d^2 \Gamma \over dq^2 d z}dz
\over\displaystyle \int_0^1{d^2 \Gamma \over dq^2 d z}dz
+\int^0_{-1}{d^2 \Gamma \over dq^2 d z}dz} \; .
 \label{asim_def}
\end{equation}
 We
show in Fig. 2 the
 predictions for the SM,  $1/R=250$ GeV  and
$ 1/R=200$ GeV. The zero of ${\cal A}_{fb}$ is  sensitive
 to the compactification parameter, so that  its experimental determination  would  constrain $1/R$.
At present, the analysis performed by Belle Collaboration
indicates that the relative sign of  $C_9$ and $C_7$ is negative,
confirming that ${\cal A}_{fb}$ should have a zero
\cite{Abe:2005km}.
\begin{figure}[htb]
\vspace{-0.3cm}\begin{center}
\hspace{-0.3cm}\psfig{file=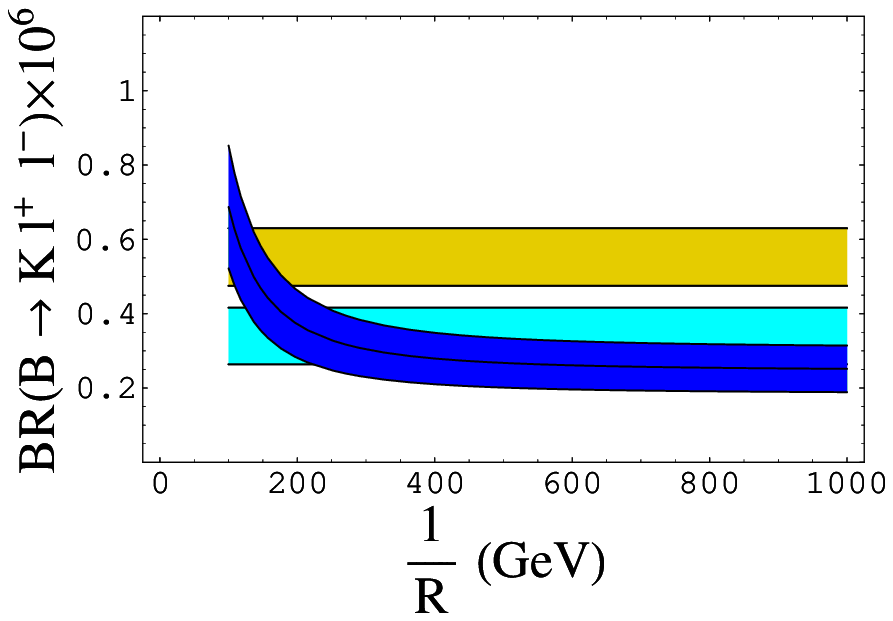,
width=0.244\textwidth}\psfig{file=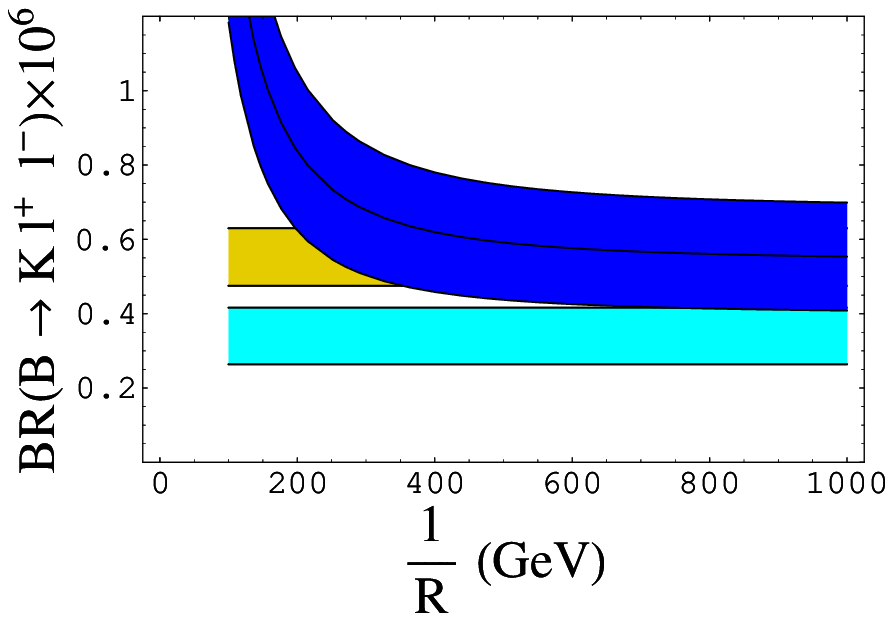,
width=0.244\textwidth}\hspace{-0.2cm}\\\hspace{-0.3cm}\psfig{file=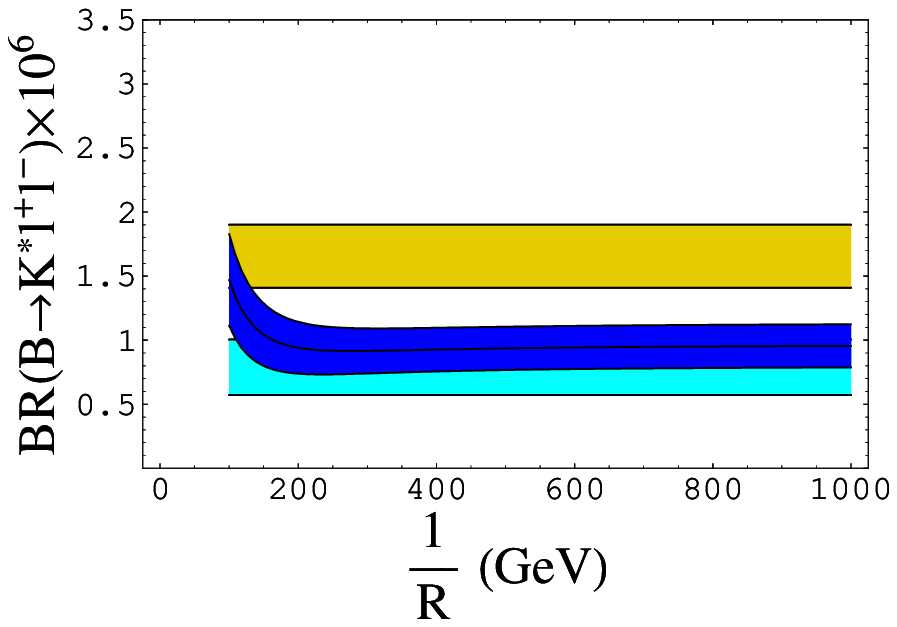,
width=0.244\textwidth}\psfig{file=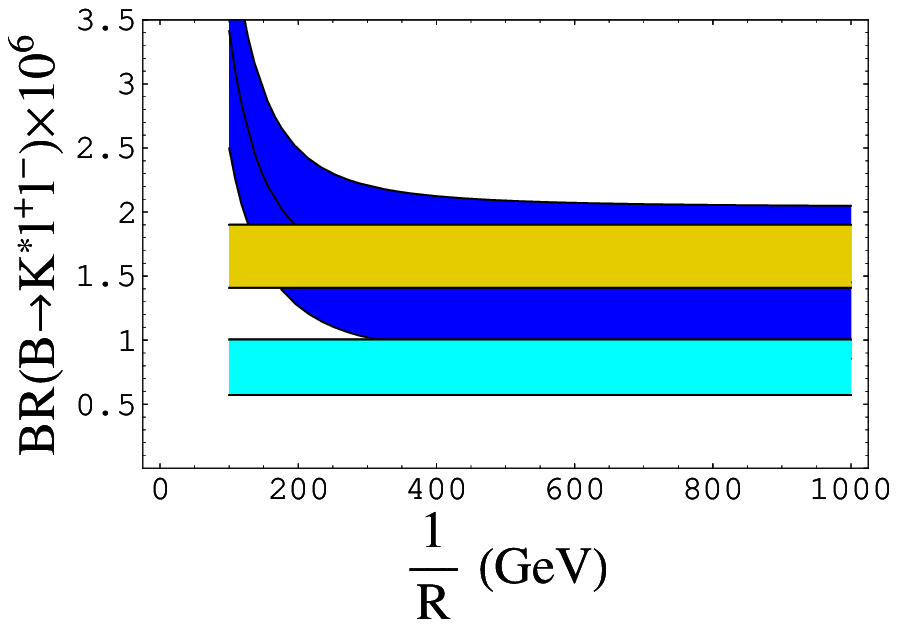,
width=0.244\textwidth}\hspace{-0.2cm} \vspace{-1.cm}\caption{$BR(B
\to K \ell^+ \ell^-)$ (upper figures) and $BR(B \to K^* \ell^+
\ell^-)$ (lower figures)  versus $1/R$ using set A (left) and B
 (right) of form factors.  The two horizontal regions refer
to  BaBar (lower band) and Belle (upper band) data.}
\end{center}\vspace{-1cm}\label{brkcol}
\end{figure}
%
\begin{figure}[htb]
\begin{center}
\hspace{-0.3cm}\psfig{file=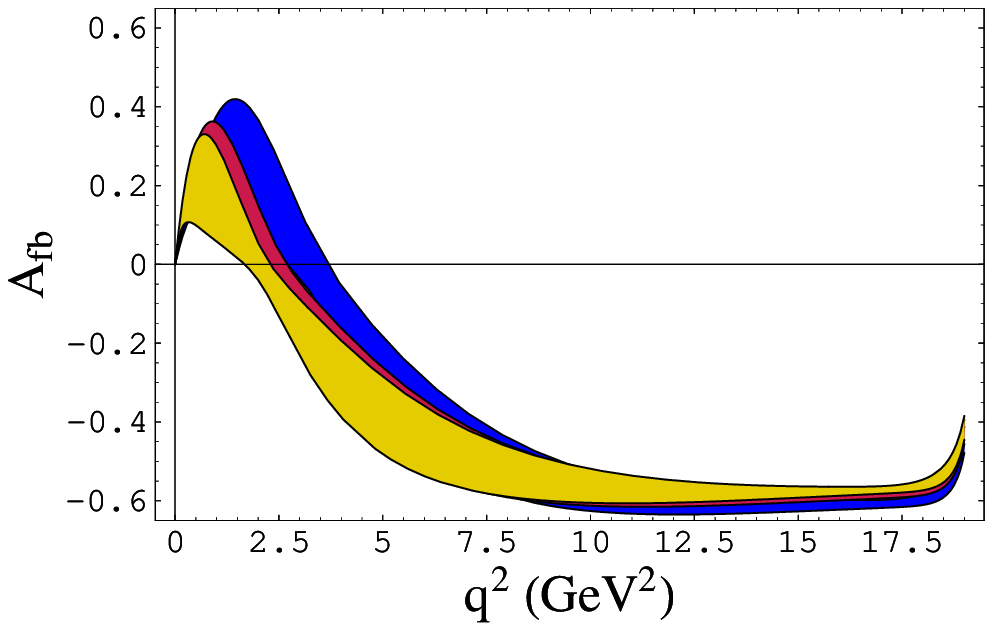,
width=0.23\textwidth}\psfig{file=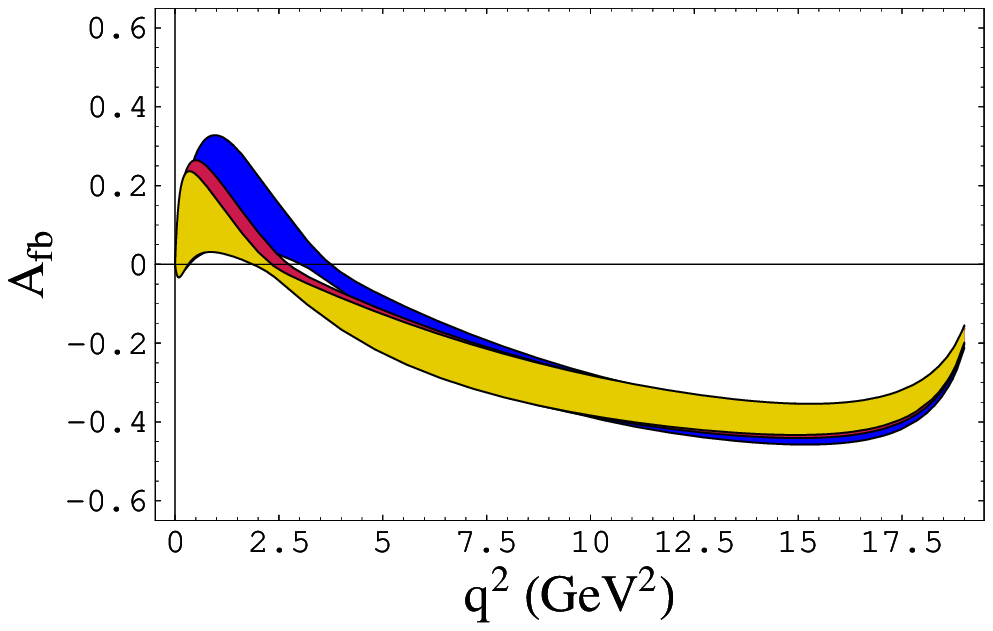,
width=0.23\textwidth}\hspace{-0.2cm}\vspace{-0.8cm}\caption{Forward-backward
lepton asymmetry in $B \to K^* \ell^+ \ell^-$  versus $1/R$ using
set A  (left) and B (right). The  dark (blue) bands correspond to
the SM results, the intermediate (red) band to $1/R=250$ GeV, the
light (yellow) one to
 $1/R=200$ GeV.}
\end{center}\vspace{-1cm}\label{fig-afb}
\end{figure}
%

We considered also the case of the modes $B \to K^{(*)} \tau^+
\tau^-$, i.e. with a massive lepton. These modes have not been
observed yet, so that it is not possible to compare their
branching ratios with data. However their analysis in the ACD
model shows that an eventual measurement of a branching ratio
larger than $2 \cdot 10^{-7}$ would be incompatible with the SM,
independently of the set of form factors used.

It is also interesting to consider the asymmetry in the $\tau^-$
polarization, defined as: \be {\cal A}_A(q^2)=\displaystyle{ {d
\Gamma \over dq^2}(s_A)-{d \Gamma \over dq^2}(-s_A) \over {d
\Gamma \over dq^2}(s_A)+{d \Gamma \over dq^2}(-s_A) }
\label{def-pol} \ee with $A=L,T,N$ and $s_L={1 \over m_\tau}(|\vec
k_1|,0,0,k_1^0)$, $s_T=(0,0,-1,0)$, $s_N=(0,1,0,0)$ being the
$\tau^-$ longitudinal, transverse and normal polarization vectors,
and $k_1$ its momentum in the lepton pair rest frame. The
longitudinal asymmetry ${\cal A}_L$ shows a mild dependence on
$1/R$, while the transverse asymmetry ${\cal A}_T$,  is a more
sensitive observable as displayed in Fig. 3 for the representative
case of $B \to K \tau^- \tau^+$.
\begin{figure}[htb]
\begin{center}
\hspace{-0.3cm}\psfig{file=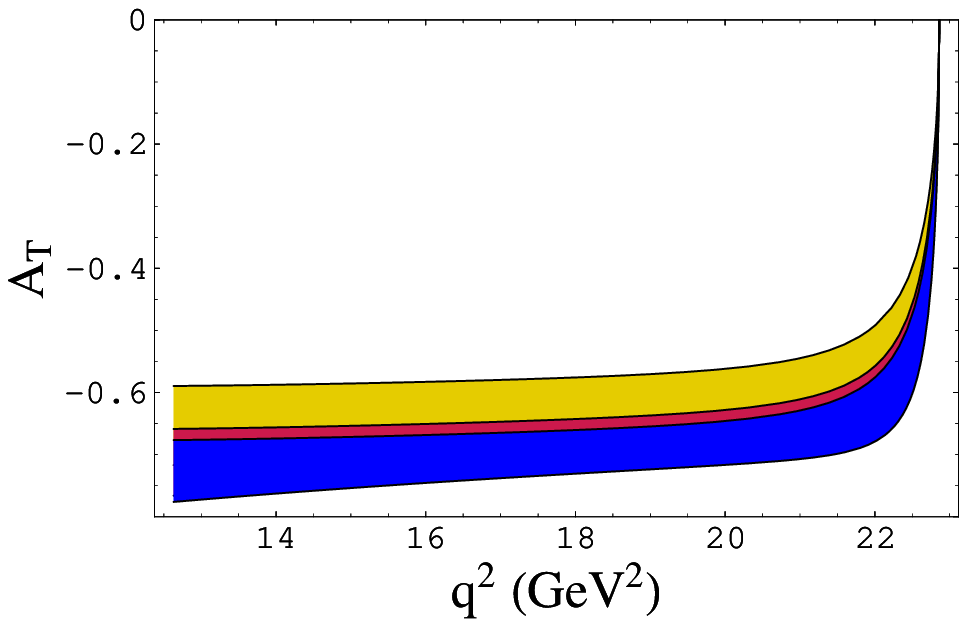,
width=0.23\textwidth}\psfig{file=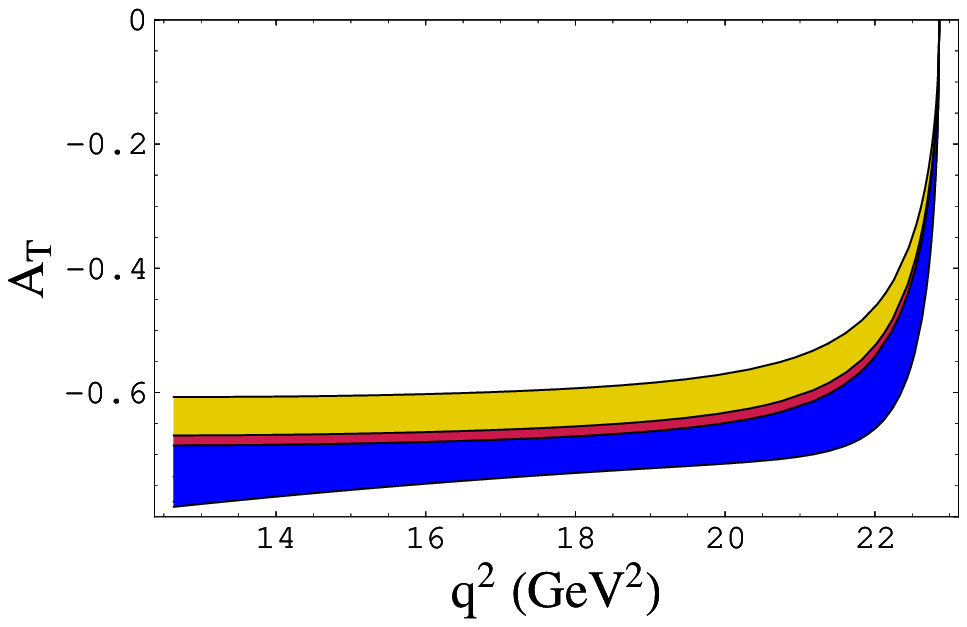,
width=0.23\textwidth}\hspace{-0.2cm}\vspace{-0.8cm}\caption{Transverse
 $\tau^-$ polarization asymmetry in $B \to K \tau^+ \tau^-$
 obtained using  set A   (left) and B  (right) of form factors.
 The  dark (blue) region is obtained in  SM,
 the intermediate (red) one for $1/R=500$ GeV, the light (yellow) one for $1/R=200$ GeV.}
\end{center}\vspace{-1cm}\label{asymTk}
\end{figure}
%
\section{THE DECAYS $B \to K^{(*)} \nu {\bar \nu}$}

 In the
SM the effective Hamiltonian  governing    $ b \to s \nu {\bar
\nu} $  induced decays is  \bea {\cal H}_{eff} &=&
\displaystyle{G_F \over \sqrt 2} {\alpha \over 2 \pi
\sin^2(\theta_W)} \; V_{ts} V^*_{tb} \; \eta_X X(x_t)\nonumber \\
&&{\bar b} \gamma^\mu ( 1- \gamma_5) s \; {\bar \nu} \gamma_\mu (
1- \gamma_5) \nu \label{hamilnunubar} \eea obtained from $Z^0$
penguin and box diagrams dominated by
 the intermediate top quark. In
(\ref{hamilnunubar})  $\theta_W$ is the Weinberg angle. The
function $X$ was computed in \cite{inami,buchalla} in the SM and
in the ACD model in  \cite{Buras:2002ej,Buras:2003mk}. We put to
unity the QCD factor $\eta_X$ \cite{buchalla,Buchalla:1998ba}.

$B \to K^{(*)} \nu {\bar \nu}$ decays have been studied within the
SM \cite{Colangelo:1996ay}. However,
 only an experimental upper bound exists  for  $B \to K
\nu \bar \nu$: $BR(B^- \to K^- \nu \bar \nu)  < 3.6 \times 10^{-5}
\,\,(90 \% \,\, CL)$ \cite{Abe:2005bq},  $BR(B^- \to K^- \nu \bar
\nu)<5.2 \times 10^{-5} \,\,\,\,(90 \% \,\, CL)$
\cite{Aubert:2004ws}. Furthermore the $1/R$ dependence, studied in
\cite{Colangelo:2006vm} turns out to be  too mild for
distinguishing values above $1/R \ge 200$ GeV.

\section{THE DECAY $ B \to K^* \gamma$ }
The transition $b \to s \gamma$ is described  by the operator
$O_7$. The most recent measurements for the exclusive branching
fractions are provided by Belle \cite{Nakao:2004th}: $BR(B^0 \to
K^{*0} \gamma)=(4.01 \pm 0.21 \pm 0.17 ) \times 10^{-5}$, $BR(B^-
\to K^{*-} \gamma)=(4.25 \pm 0.31 \pm 0.24) \times 10^{-5}$ and
BaBar \cite{Aubert:2004te}: $BR(B^0 \to K^{*0} \gamma)=(3.92 \pm
0.20\pm 0.24 )
 \times 10^{-5}$, $BR(B^- \to K^{*-} \gamma)=(3.87 \pm 0.28 \pm 0.26 ) \times 10^{-5}$.

 In Fig. 4 the
 branching ratio computed in the ACD model
 is plotted versus  $1/R$: the sensitivity to the
 this
 parameter is evident;  a lower bound of
  $1/R \ge 250$ GeV  can be put adopting set A, and a stronger bound of    $1/R \ge
400$ GeV using set B, which is the most stringent bound that can
be currently put on this parameter from the  $B$ decay modes we
have considered.
\begin{figure}[htb]
\vspace{-0.5cm}\begin{center}
\hspace{-0.2cm}\psfig{file=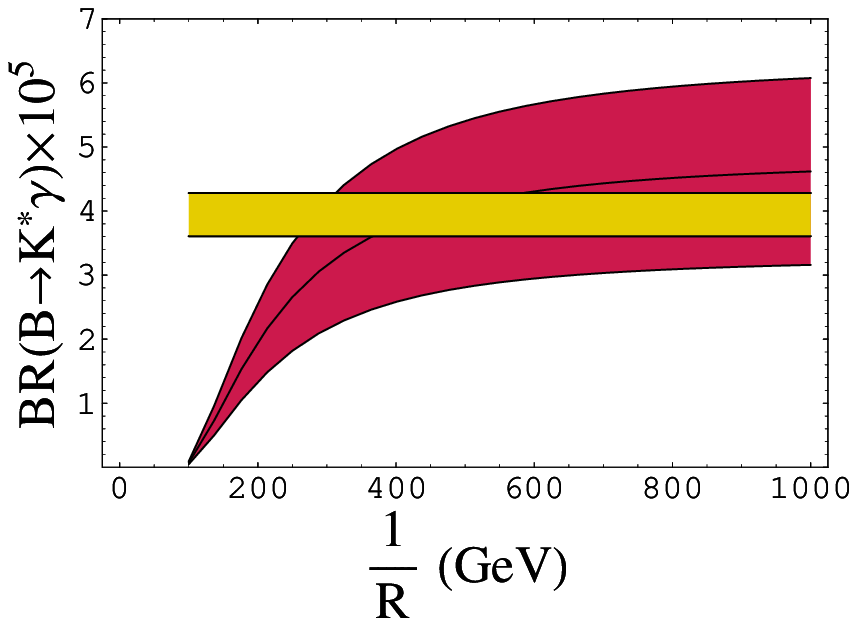,
width=0.25\textwidth}\psfig{file=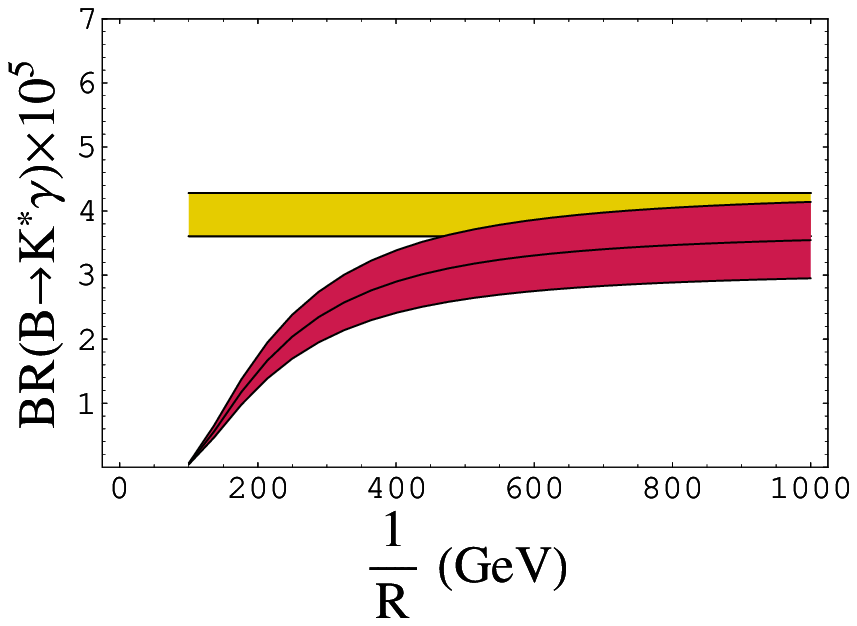,
width=0.25\textwidth}\hspace{-0.2cm} \vspace{-1.cm}\caption{$BR(B
\to K^* \gamma)$  versus $1/R$ using  set A  (left) and   B
 (right) of form factors .  The  horizontal
band corresponds to the experimental result.}
\end{center}\vspace{-1cm}\label{brkstargamma}
\end{figure}
%

\section{CONCLUSIONS AND PERSPECTIVES} \label{sec:concl}
We have shown how  the predictions for    $B \to K^{(*)} \ell^+
\ell^-$, $B \to K^{(*)}\nu \bar \nu$, $B \to K^* \gamma$ decays
are modified within the ACD scenario. The constraints on  $1/R$
are
  slightly model dependent,   being
different using different sets of form factors. Nevertheless,
 various distributions, together with the lepton
forward-backward asymmetry in $B \to K^* \ell^+ \ell^-$ are very
promising in order to constrain $1/R$,  the most stringent lower
bound coming from $B \to K^* \gamma$. Improvements in the
experimental data, expected in the near future, will allow to
establish more stringent constraints for the compactification
radius.

\end{document}